\documentstyle[epsf]{mn}

\title[]{{\sl XMM-Newton} observations of three short period polars:
V347 Pav, GG Leo and EU UMa}

\author[Ramsay et al]{
Gavin Ramsay$^{1}$, Mark Cropper$^{1}$, K. O. Mason$^{1}$, F. A.
C\'{o}rdova$^{2}$, W. Priedhorsky$^{3}$\\
$^{1}$Mullard Space Science Laboratory, University College London,
Holmbury St. Mary, Dorking, Surrey, RH5 6NT, UK\\
$^{2}$University of California, Riverside, CA 92521, USA\\
$^{3}$Los Alamos National Laboratory, MS D436, Los Alamos, NM 87545, USA}
\date{Received: }

\begin{document}
\outer\def\gtae {$\buildrel {\lower3pt\hbox{$>$}} \over 
{\lower2pt\hbox{$\sim$}} $}
\outer\def\ltae {$\buildrel {\lower3pt\hbox{$<$}} \over 
{\lower2pt\hbox{$\sim$}} $}
\newcommand{\ergscm} {ergs s$^{-1}$ cm$^{-2}$}
\newcommand{\ergss} {ergs s$^{-1}$}
\newcommand{\ergsd} {ergs s$^{-1}$ $d^{2}_{100}$}
\newcommand{\pcmsq} {cm$^{-2}$}
\newcommand{\ros} {\sl ROSAT}
\newcommand{\exo} {\sl EXOSAT}
\newcommand{\xmm} {\sl XMM-Newton}
\def\rchi{{${\chi}_{\nu}^{2}$}}
\def\uchi{{${\chi}^{2}$}}
\newcommand{\Msun} {$M_{\odot}$}
\newcommand{\Mwd} {$M_{wd}$}
\def\Mdot{\hbox{$\dot M$}}
\def\mdot{\hbox{$\dot m$}}

\maketitle

\begin{abstract}

We present phase-resolved {\xmm} data of three short period polars:
V347 Pav, GG Leo and EU UMa. All three systems show one dominant
accretion region which is seen for approximately half of the orbital
cycle. GG Leo shows a strong dip feature in its X-ray and UV light
curves which is due to absorption of X-rays from the accretion site by
the accretion stream. The emission in the case of EU UMa is dominated
by soft X-rays: its soft/hard X-ray ratio is amongst the highest seen
in these objects. In contrast, GG Leo and V347 Pav shows a ratio
consistent with that predicted by the standard shock model. We infer
the mass of the white dwarf and explore the affect of restricting the
energy range on the derived parameters.

\end{abstract}


\section{Introduction}

Polars or AM Her systems are accreting binary systems in which
material transfers from a dwarf secondary star onto a magnetic
($B\sim$10--200MG) white dwarf through Roche lobe overflow. At some
height above the photosphere of the white dwarf a shock forms. Hard
X-rays are generated in this post-shock flow. Some of these X-rays are
intercepted by the photosphere of the white dwarf and are re-emitted
at lower energies. Soft X-rays can also be produced by dense `blobs'
of material which impact directly into the white dwarf. Further,
cyclotron emission in the optical band is produced by electrons
spiralling around the magnetic field lines (see Warner 1995 for a
review).

To better characterise the X-ray emission from polars we have
undertaken a survey of 37 systems using {\xmm}. An overview of the
survey and initial results are given in Ramsay \& Cropper (2003b). We
have presented phase-resolved observations of DP Leo, WW Hor (Ramsay
et al 2001), BY Cam (Ramsay \& Cropper 2002a), CE Gru (Ramsay \&
Cropper 2002b), EV UMa, RX J1002--1925 and RX J1007--2016 (Ramsay \&
Cropper 2003a). Here, we present phase-resolved observations on three
further polars, V347 Pav (RE J1844--741), GG Leo (RX J1015.6+0904) and
EU UMa (RE J1149+28), all of which were discovered using {\ros} and
have short periods (90, 80 and 90 mins respectively).

\section{Present and Past Observations}

{\xmm} was launched in Dec 1999 by the European Space Agency. It has
the largest effective area of any X-ray satellite and also has a 30 cm
optical/UV telescope (the Optical Monitor, OM: Mason et al 2001)
allowing simultaneous X-ray and optical/UV coverage. The EPIC
instruments contain imaging detectors spanning the energy range
0.1--10keV with moderate spectra resolution. The OM data were taken in
two UV filters (UVW1: 2400--3400 \AA, UVW2: 1800--2400 \AA) and one
optical band ($V$ band). The observation log is shown in Table
\ref{log}. The data obtained using the RGS instruments were of low
signal-to-noise and are therefore not discussed further. The data were
processed using the {\sl XMM-Newton} {\sl Science Analysis Software}
v5.3.3. For details of the analysis procedure see Ramsay \& Cropper
(2003a). We show in Table \ref{log} the mean $V$ mag determined using
the OM for each source at the time of our observation.

V347 Pav was discovered using the {\ros} extreme UV WFC survey (Pounds
et al 1993) and identified with a $V$=16 mag Cataclysmic Variable (CV)
by O'Donoghue et al (1993). Bailey et al (1995) confirmed V347 Pav as
a polar when large amplitude circular polarisation variations were
detected with a period of 90 min. Based on a typical polar X-ray
spectrum in the high accretion state, V347 Pav was brighter during the
{\xmm} observations compared to March 1993 but not as bright as in
July 1994 (Ramsay et al 1996). It was therefore in a high accretion
state.

GG Leo was also discovered as part of the {\ros} all sky survey, on
this occasion using the X-ray Telescope. Followup observations by
Burwitz et al (1998) identified the optical counterpart as a $V=16-17$
mag polar with an orbital period close to 90 min. Burwitz et al (1998)
reported X-ray data taken from 3 epochs: it was bright on each
occasion and showed a narrow dip which was taken to be the accretion
stream obscuring the accretion region. Comparing the count rates we
find that GG Leo was at a similar X-ray brightness during these {\xmm}
observations as in May 1994, so that it was also in a high accretion
state.

EU UMa was discovered using the {\ros} extreme UV WFC survey and
identified with a $V$=17 mag CV by Mittaz et al (1992).  Observations
using {\sl EUVE} by Howell et al (1995) found that the orbital period
was likely to be 90 mins. X-ray observations using {\ros} data (Ramsay
et al 1994, Ramsay 1995) showed it had a peak count rate of $\sim$5
ct/s. We estimate that it was fainter by a factor of 3 in these {\xmm}
observations, although it is still likely to have been in a relatively
high accretion state.

\begin{table*}
\begin{center}
\begin{tabular}{lrrr}
\hline
 & V347 Pav & GG Leo & EU UMa\\
\hline
Date & 2002 Mar 16 & 2002 May 13 & 2002 Jun 12\\
EPIC MOS & Small Window thin 5980 sec & Small Window thin 7975 sec &
 Small Window thin 5792 sec\\
EPIC pn & Small Window thin 5053 sec& Small Window thin 7053 sec&
 Small Window thin 4870 sec\\
RGS & 6453 sec & 8448 sec & 6265 sec\\
OM & Image/fast UVW1 1500 sec & Image/fast UVW1 2500 sec &Image/fast
 UVW1 1500 sec\\
OM & Image/fast UVW2 2000 sec& Image/fast UVW2 3000 sec &Image/fast
 UVW2 2000 sec \\
OM & Image/fast V 1500 sec& Image/fast V 1900 sec& Image/fast V 1500 sec\\
Mean $V$ mag & 16.8 & 17.1 & 18.2 \\
Accretion state & High & High & High\\
\hline
\end{tabular}
\end{center}
\caption{The log of {\xmm} observations of V347 Pav, GG Leo \& EU UMa.
`Thin' refers to the filter used. The exposure time in each detector
is shown in seconds. The UVW1 filter has a coverage 2400--3400
\AA\hspace{1mm} and UVW2 1800--2400 \AA.}
\label{log}
\end{table*}

\section{Light curves}

\subsection{V347 Pav}
\label{v347}

\begin{figure}
\begin{center}
\setlength{\unitlength}{1cm}
\begin{picture}(8,10.5)
\put(-0.5,-1.8){\includegraphics{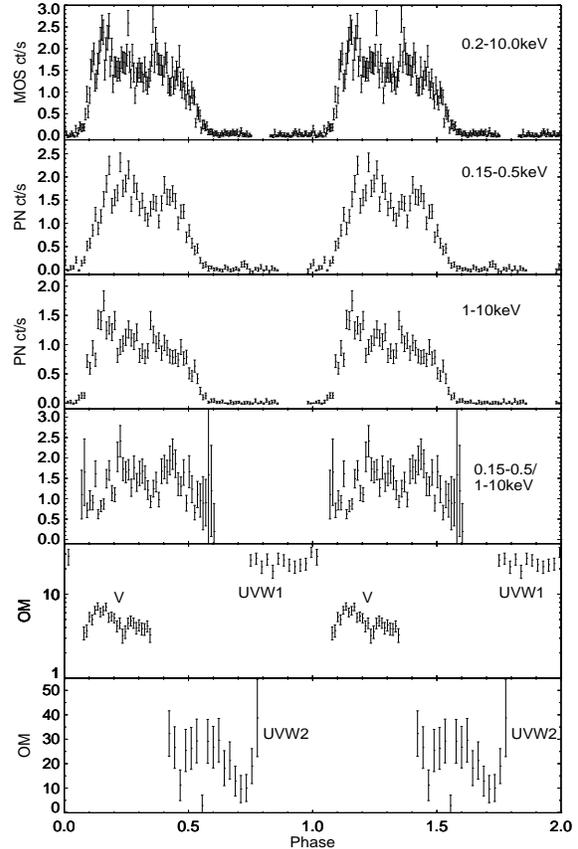}}
\end{picture}
\end{center}
\caption{The phased and binned X-ray and optical light curves of V347
Pav.  We have folded the data on the ephemeris reported in \S \ref{v347}.
Energy bands 0.15--0.5keV and 1.0--10.0keV have bin widths of 60 sec
(=0.011 cycles), while the energy band 0.2-10.0keV has a bin width of
30 sec (=0.0055 cycles). The optical data were binned into 60sec bins,
while the UV data was binned into 120 sec bins (=0.022 cycles). The
units for the OM data are $10^{-16}$ \ergscm \AA$^{-1}$.}
\label{lightv347} 
\end{figure}

V347 Pav is relatively bright in soft and hard X-rays, peaking at over
1 ct/s (in the EPIC pn detector) in both soft and hard X-ray bands. It
shows a distinct faint and bright phase lasting nearly 0.5 cycles each
(Figure \ref{lightv347}). The count rate in the faint phase is
significant at 0.037$\pm$0.007 ct/s. The intensity shows a rapid rise
at $\phi$=0.0 when the bright accretion region comes into view over
the limb of the white dwarf. As seen from the hardness ratio curve
(Figure \ref{lightv347}), the hard X-ray light curve rises more
rapidly than the soft X-rays. This is expected since in the standard
models (eg Lamb \& Masters 1979, King \& Lasota 1979) hard X-rays are
optically thin whereas the soft X-rays are optically thick. The rapid
but not instantaneous rise in the hard X-ray flux indicates that the
shock has a significant vertical height above the photosphere of the
white dwarf and/or the accretion region is extended.  Based on our
emission model described in \S \ref{model} we predict that for a white
dwarf mass of 1.0\Msun, a specific accretion rate of 1 g$^{-1}$
s$^{-1}$ cm$^{-2}$ and a magnetic field strength of 20MG (Potter,
Cropper \& Hakala 2000), the shock height would be $\sim$0.1
$R_{wd}$. This is high enough to cause the observed difference in the
rise time in the hard X-ray light curve. From the hardness ratio plot,
the light curve is harder near the center of the bright phase
($\phi\sim$0.35) indicating that some absorption affect maybe present
at these phases. The $V$ band data show a rise in flux at the same
phase as the onset of the bright phase.

Phase zero on the optical ephemeris of Ramsay et al (1996) was defined
as the start of the rapid rise from the faint phase. The accumulated
error in that ephemeris is $\phi$=0.02. Phasing our data on that
ephemeris shows the bright phase starting at a much earlier phase
($\phi$=0.8). We revisited the optical data and included the timing of
the start of the bright phase reported by Bailey et al (1995). We find
that the period of Ramsay et al (1996) was actually a one day alias of
the true orbital period. The best fit ephemeris is:

\begin{equation}
T = HJD 2448475.2913(5) + 0.062557097(36)
\end{equation}

We have therefore used the above ephemeris to phase our data. We also
re-analysed the {\ros} data shown by Ramsay et al (1996). We find that
all those data are consistent with the main accretion region being the
prime source of X-rays. The fact that Ramsay et al (1996) found the
phase of the bright X-ray region to vary is phase was therefore due to
the use of an incorrect ephemeris.

\subsection{GG Leo}

\begin{figure}
\begin{center}
\setlength{\unitlength}{1cm}
\begin{picture}(8,9.5)
\put(0.5,-1.5){\includegraphics{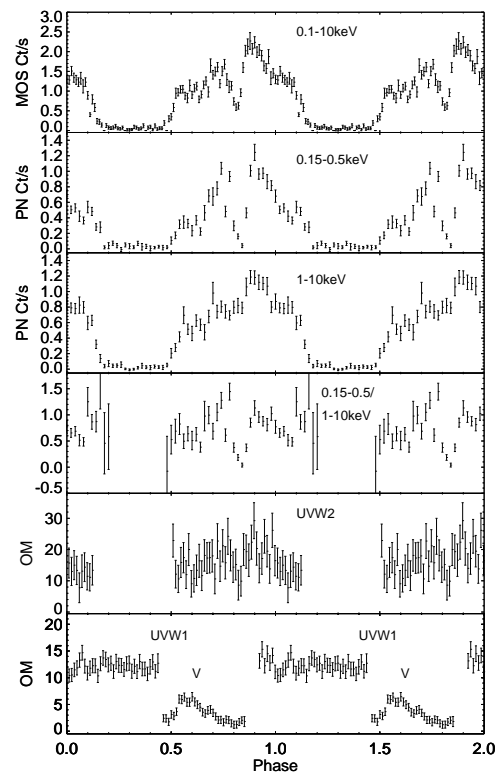}}
\end{picture}
\end{center}
\caption{The binned X-ray and OM light curves of GG Leo. We fold the
data on the orbital period of Burwitz et al (1998). The phasing is
arbitrary. The 0.15--1.0keV data have been binned into 0.01 cycle bins
while the 1.0--10.0keV band data and the OM data have been binned into
0.02 cycle bins. The units for the OM data are $10^{-16}$ \ergscm \AA$^{-1}$.}
\label{lightrx1002} 
\end{figure}

Like V347 Pav, GG Leo shows a relatively high count rate in both X-ray
bands ($\sim$1 ct/s in the EPIC pn detector). It has a faint phase
lasting $\Delta\phi\sim0.3$ cycles and a bright phase lasting
$\Delta\phi\sim0.7$ (Figure \ref{lightrx1002}). This implies the X-ray
bright accretion region is located in the upper hemisphere of the
white dwarf or is significantly extended. Although faint, the source
is again clearly visible during the faint phase (0.050$\pm$0.009
ct/s). This soft X-ray light curve is similar to those seen using
{\ros} data (Burwitz et al 1998). Burwitz et al (1998) also report
$RI$ band light curves which are similar to our $V$ band data and are
consistent with significant cyclotron emission from the main accretion
region.

There is a deep dip in soft energies during the bright phase
($\phi=$0.84 in Figure \ref{lightrx1002}). These dips are due to the
accretion stream obscuring the accretion region on the white dwarf
({\sl cf} Watson et al 1989). A similar dip is also seen in HST UV
data of the polar QS Tel (de Martino et al 1998). Using the fit to the
bright phase spectrum (which excludes the dip) shown in \S \ref{spec}
as the reference model, we find that the neutral absorption must
increase to $\sim2\times10^{21}$ \pcmsq to obtain the observed count
rate in the 0.15--0.5keV energy band. This level of absorption has
negligible affect on the count rate in the 1-10keV band.

To estimate the amount of absorption required to produce the decrease
in the flux in the UVW2 band we used the T\"{u}bingen-Boulder
inter-stellar medium absorption model. (Other absorption models
included in {\tt XSPEC} are not sensitive to absorption in the
UV). This includes gas, grain and $H_{2}$ absorption and recent
improvements in the photo-ionisation cross-sections: since we do not
expect dust or $H_{2}$ molecules to be present in polars, we switch
these parameters to zero. We find that the equivalent Hydrogen column
density required to reduce the UVW2 flux by the observed amount is
7--8$\times10^{21}$ \pcmsq. This amount of absorption would have a
noticeable affect on the 1--10keV flux, which we do not observe. The
T\"{u}bingen-Boulder model is for a neutral absorber, and depending on
the location of the absorber it will be irradiated to some degree. The
fact that the result from this fit is inconsistent with the X-rays is
indicative of such irradiation and/or other complex processes.

\subsection{EU UMa}

\begin{figure}
\begin{center}
\setlength{\unitlength}{1cm}
\begin{picture}(8,9)
\put(-0.5,-1.4){\includegraphics{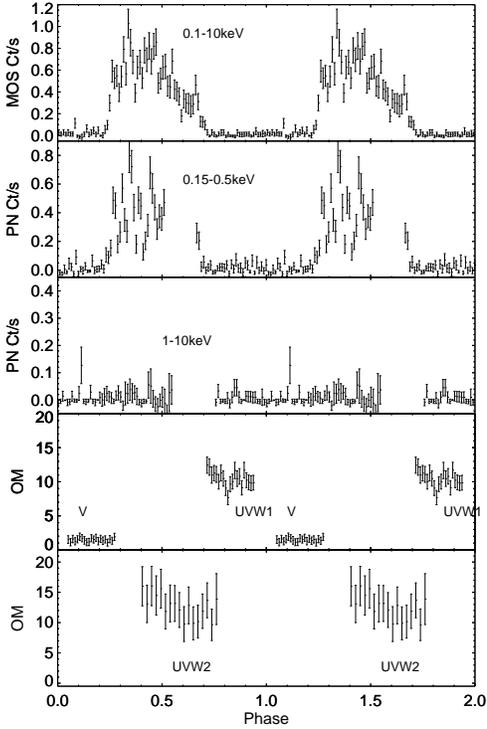}}
\end{picture}
\end{center}
\caption{The X-ray and OM light curves of EU UMa. We fold the data on
the orbital period of Howell et al (1995). We define the folding epoch
(arbitrarily) at HJD=2452000. The 0.15--1.0keV data has been binned
into 0.01 cycle bins and the 1.0--10.0keV data into 0.05 cycle
bins. The UVW1 and V band OM data has been binned into 60 sec bins
(=0.011 cycles) and the UVW2 into 120 sec bins (=0.022 cycles).
The units for the OM data are $10^{-16}$ \ergscm \AA$^{-1}$.}
\label{lighteuuma} 
\end{figure}

Like V347 Pav and GG Leo, EU UMa is principally a one pole system,
showing a bright phase lasting $\sim$0.5 cycles (Figure
\ref{lighteuuma}). However, in contrast to V347 Pav and GG Leo, it
shows very little flux above 1keV. This is similar to the {\ros} X-ray
observations (Ramsay 1995). During the faint phase the source is
weakly detected 0.012$\pm$0.004 ct/s in the EPIC pn detector.

Again in contrast to V347 Pav and GG Leo, the soft X-ray band shows
prominent flaring activity.  We performed a Discrete Fourier Transform
on the soft X-ray light curve, but there was no evidence that the
flare activity was modulated on a quasi-coherent period.

The OM data shows no large variations in the flux levels over the
orbital period. The $V$ band data cover the same phase interval as the
sharp rise in X-ray flux. Unusually, there is no corresponding rise in
the optical flux. Mittaz et al (1992) note that EU UMa shows an
unusually high EUV/optical ratio. The UVW1 data were taken in the
faint phase and show no significant variation, while there is some
evidence that the UVW2 data reflect the decline from the bright
phase.

\section{X-ray spectra}
\label{spec}

We extracted X-ray spectra from the bright phase of the 3 systems. In
the case of GG Leo we excluded the phase interval where the prominent
dip feature was seen. We extracted single and double events and used
the response matrix epn\_sw20\_sdY9.rmf for the EPIC pn data together
with an auxiliary file generated using the SAS task {\tt arfgen}. We
also extracted spectra using the EPIC MOS spectra. The results
obtained using the MOS spectra were consistent with the pn spectra,
but we show only the results using the pn data (for the sake of
brevity) since these had the higher signal to noise ratio.

\subsection{The model}
\label{model}

We modelled the data using a simple neutral absorber and an emission
model of the kind described by Cropper et al (1999). We have used this
model in our previous studies of polars observed using {\xmm} - we
refer the reader to those papers for details (see \S 1). We fix the
specific accretion rate at 1 g s$^{-1}$ cm$^{-2}$ and the ratio of
cooling due to cyclotron to bremsstrahlung at the shock at 5. We also
added a blackbody component and, if necessary, a neutral absorber with
a partial covering fraction. We show in Table \ref{fits} the derived
spectral parameters with the goodness of fits.

\begin{table*}
\begin{center}
\begin{tabular}{llrrrrrrrr}
\hline 
Source & $N_{H}$ & $kT_{bb}$ & $M_{wd}$ &
Blackbody& X-hard & $L_{bb,bol}$ &  $L_{X-hard,bol}$ & $L_{bb,bol}/$ & 
\rchi \\
       & ($10^{20}$) & (eV) & (\Msun) &
Flux & Flux & &  & $L_{X-hard,bol}$ & (dof) \\
       &  \pcmsq        &      &  &($\times10^{-11}$)
& ($\times10^{-11}$) & ($\times10^{31}$) & ($\times10^{31}$) & 
($\times10^{31}$) &  \\
\hline
EU UMa & 0.1$^{+3.7}_{-0.1}$ &  20$^{+3}_{-7}$ & 1.2$^{+0.2}_{-0.1}$
& 12.7$_{-7.5}^{+17600}$ & 
0.06$^{+0.02}_{-0.05}$ & 3.8$^{+5300}_{-2.3}$ &
0.08$^{+0.03}_{-0.06}$ & 46.9$_{-33.3}^{+2780}$ & 0.95 (28)\\
GG Leo & 0.0$^{+0.6}$ & 
38$^{+4}_{-3}$ & 1.13$\pm$0.03 &  
2.0$_{-0.2}^{+0.7}$ & 3.10$^{+0.06}_{-0.13}$ & 0.6$_{-0.1}^{+0.2}$ & 
3.7$_{-0.2}^{+0.1}$ & 0.16$^{+0.07}_{-0.02}$ & 1.40 (136) \\
 & 840$^{+220}_{-160}$, 0.65$\pm0.02$ & & & & & & &  \\
V347 Pav & 0.0$^{+0.4}$ & 32$\pm3$
& 1.00$\pm$0.06 & 
6.4$^{+2.2}_{-0.4}$ & 2.21$_{-0.09}^{+0.08}$  & 1.9$_{-0.1}^{+0.6}$ & 
2.7$^{+0.1}_{-0.2}$ & 0.70$^{+0.3}_{-0.1}$ & 1.02 (237) \\
 & 5.6$^{+3.6}_{-1.6}$ 0.50$\pm$0.03 & & & & & & &  \\
\hline
\end{tabular}
\end{center}
\caption{The fits to the X-ray data.  The blackbody bolometric
luminosity, $L_{bb,bol}$ and the X-ray luminosity from the shock,
$L_{Xhard-bol}$, are defined in the text. The units of flux are
\ergscm and luminosity \ergss: we assume a distance of 100 pc in each
case. The metal abundance was fixed at solar. The errors are quoted at
the 90 percent confidence level.}
\label{fits}
\end{table*}

\subsection{The spectral fits}

\begin{figure}
\begin{center}
\setlength{\unitlength}{1cm}
\begin{picture}(8,5.)
\put(-0.5,-0.7){\includegraphics{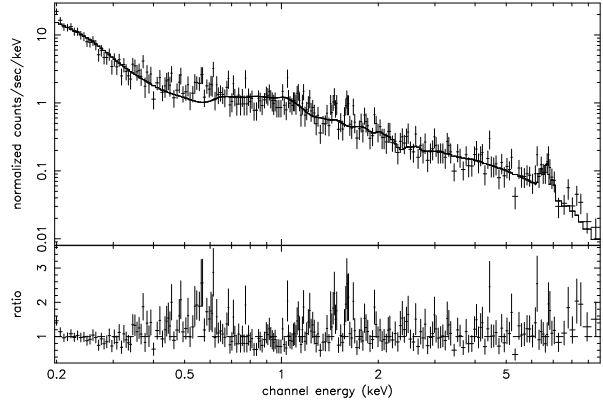}}
\end{picture}
\end{center}
\caption{The bright phase EPIC pn spectra of V347 Pav. The best fit
model is shown as a solid line.}
\label{specv347pav} 
\end{figure}

\begin{figure}
\begin{center}
\setlength{\unitlength}{1cm}
\begin{picture}(8,4.8)
\put(-0.5,-0.7){\includegraphics{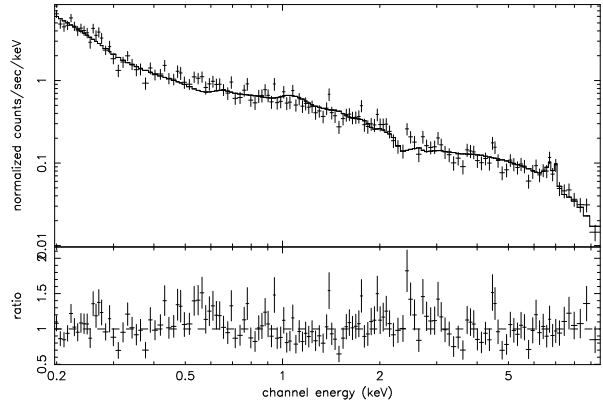}}
\end{picture}
\end{center}
\caption{The bright phase EPIC pn spectra of GG Leo (the accretion
stream dip interval has been excluded). The best fit model is shown as
a solid line.}
\label{specggleo} 
\end{figure}

\begin{figure}
\begin{center}
\setlength{\unitlength}{1cm}
\begin{picture}(8,5)
\put(-0.5,-0.7){\includegraphics{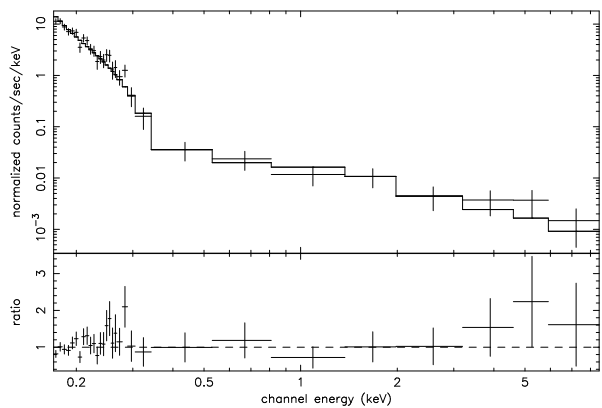}}
\end{picture}
\end{center}
\caption{The bright phase EPIC pn spectra of EU UMa. The best fit
model is shown as a solid line.}
\label{speceuuma} 
\end{figure}

We show the EPIC pn spectra in Figure \ref{specv347pav} (V347 Pav),
Figure \ref{specggleo} (GG Leo) and Figure \ref{speceuuma} (EU
UMa). As expected from the energy resolved light curves, both GG Leo
and V347 Pav show strong hard X-ray components, while this is weak in
EU UMa. We obtain good fits (Table \ref{fits}) to the spectra of EU
UMa and V347 Pav, with significant residuals in GG Leo
(\rchi=1.40). In the case of GG Leo and V347 Pav we require the
addition of a neutral absorber with partial covering to achieve good
fits.  In the case of V347 Pav (and to a lesser extent GG Leo) there
are prominent residuals near 0.5--0.6keV and also 1.45 and
1.60keV. Our model includes line emission from collisionally ionised
plasma but does not include emission from photo-ionised plasma, seen
in sources such as X-ray binaries where there is a strong irradiating
source. One prominent photo-ionised emission line is the OVII He-like
triplet near 0.57 keV which may be the source of the residuals at
these energies.  Emission lines near 1.45keV and 1.60keV are Mg XII
and Mg XI.

\subsection{The energy balance}

In Table \ref{fits} we show the luminosities of both the soft and hard
X-ray components. We define the hard X-ray luminosity as
$L_{X-hard,bol}=4\pi$Flux$_{X-hard,bol}d^{2}$ where
Flux$_{X-hard,bol}$ is the unabsorbed, bolometric flux from the hard
X-ray component and $d$ is the distance and the soft X-ray luminosity
as $L_{soft,bol}=\pi$Flux$_{soft,bol}$sec($\theta)d^{2}$. We refer the
reader to Ramsay \& Cropper (2003a) for further details.

We show in Table \ref{fits} the ratio $L_{bb,bol}/L_{X-hard,bol}$ for
each source. These ratios do not include the geometrical correction
term $\sec(\theta)$ for the blackbody component. For EU UMa and GG Leo
we do not know the viewing angle to the accretion region. A mean
viewing angle of 50$^{\circ}$ and 80$^{\circ}$ implies a correction of
1.55 and 5.8 respectively. Even without a geometrical correction, EU
UMa shows one of the highest soft/hard ratios ever observed in a
polar. Its X-ray emission is likely to be dominated by the accretion
of dense blobs which impact the photosphere of the white dwarf
directly without forming a shock (Kuijpers \& Pringle 1982). In
contrast GG Leo shows an energy balance consistent with the standard
accretion model (which predicts a ratio $\sim$0.5) for a high mean
viewing angle. For V347 Pav, Potter, Cropper \& Hakala (2000) estimate
an inclination of 64--72$^{\circ}$, a dipole offset less than
8$^{\circ}$ and that the main accretion pole is located near the
magnetic equator. This implies the viewing angle is
$\sim30-40^{\circ}$ and the geometrical correction is
1.15--1.30. Therefore V347 Pav shows a small excess over the standard
shock model, but this excess is likely to disappear were the
contribution from cyclotron emission to be included. We discuss the
energy balance of our full {\xmm} sample of polars in Ramsay \&
Cropper (2003c), although a preliminary view is given in Ramsay \&
Cropper (2003b).

\subsection{The white dwarf mass}
\label{mass}

We can infer the mass of the white dwarf from our model fitting,
assuming a mass-radius relationship for the white dwarf. As usual we
assume the Nauenberg (1972) relationship. From fitting the 0.2--10keV
spectra we estimate a white dwarf mass of 1.0--1.1\Msun for all 3
polars in this study.

For sources relatively strong in hard X-rays, it is also possible to
restrict the energy range so that the fitted range is above the main
absorption edges (most of which are below $\sim$2keV). We fitted the
spectra of V347 Pav and GG Leo in the energy range 3--10keV. We
stepped through a range of white dwarf masses and metal
abundances. The fits to these spectra are shown in Table
\ref{above3kev}. For V347 Pav we find the fits are very similar for a
white dwarf mass between 0.6--1.0\Msun. For GG Leo a white dwarf of
mass 0.6\Msun gives significantly poorer fits compared to a 0.8 or
1.0\Msun white dwarf.

We then used the optimal fits to these models and fitted them over the
extended energy range 0.2-10.0keV. We kept the shock emission model
fixed and added a blackbody and partial covering model (which were
allowed to vary). The results to the fits are also shown in Table
\ref{above3kev}. We find that for V347 Pav, the resulting fit is still
reasonably good (\rchi$\sim$1.1) although the best fit model (Table
\ref{fits}) is formally a better fit at the $>$99\% confidence
level. In the case of GG Leo, for a white dwarf mass of 1.0\Msun and a
metal abundance of 0.5 solar, the fit is very similar to that found
before. Lower masses (of both solar and 0.5 solar metallicities) gave
significantly poorer fits.

Our technique of determining masses has been controversial in that for
many systems our X-ray fitting method predicts rather heavy masses
(1\Msun or more) (eg Schwope et al 2002). The main criticism has been
that for energies less than $\sim$2keV, the uncertainties in the
absorption limit the accuracy that the mass can be determined.  We
have shown here that by excluding energies less than 3keV, the masses
for GG Leo and V347 Pav were lower than when we used the full spectral
range. However, these fits were poorer at the formal level. This may
be due to the absorption being much more complex than the relatively
simple treatment assumed here. We accept that for systems which have
sufficiently strong hard X-ray components to achieve the necessary
signal to noise ratio it is desirable to use energies above 3keV to
fix this component. However, even when we do this in the case of GG
Leo the best fit mass is still $\sim$1\Msun.

We noted in \S 4.2 the presence of residuals in the spectral fits near
0.57keV which maybe due to our emission model not including
photo-ionised lines. Emission from this triplet is seen in the
intermediate polar EX Hya (Cropper et al 2002) which has a white dwarf
mass of 0.5\Msun. If the ionising source was stronger then this would
result in stronger photo-ionised lines. If the white dwarf was more
massive then the ionising source would be stronger and result in more
prominent emission near 0.57keV. This is consistent with our, rather
heavy, mass estimates.

\begin{table}
\begin{center}
\begin{tabular}{lcccccc}
\hline
Source & \multicolumn{3}{c} {3--10keV} & \multicolumn{3}{c} {0.2--10keV} \\
       & M & Z ($\odot$) & \uchi (dof) & M & Z
($\odot$) & \uchi (dof)\\
\hline
V347     & 1.0 & 1.0 & 54.3 (56) & 1.0 & 1.0 & 295.2 (242) \\
Pav      & 0.8 & 1.0 & 55.4 (56) & 0.8 & 1.0 & 268.6 (242) \\
         & 0.6 & 1.0 & 61.6 (56) & 0.6 & 1.0 & 285.6 (242) \\
         & 1.0 & 0.5 & 56.6 (56) & 1.0 & 0.5 & 271.1 (242) \\
         & 0.8 & 0.5 & 56.0 (56) & 0.8 & 0.5 & 273.5 (242) \\
         & 0.6 & 0.5 & 57.7 (56) & 0.6 & 0.5 & 290.4 (242) \\
\hline
GG       & 1.0 & 1.0 & 38.2 (35) & 1.0 & 1.0 & 221.4 (141) \\
Leo      & 0.8 & 1.0 & 47.6 (35) & 0.8 & 1.0 & 266.5 (141) \\
         & 0.6 & 1.0 & 72.1 (35) & 0.6 & 1.0 & 328.5 (141) \\
         & 1.0 & 0.5 & 37.8 (35) & 1.0 & 0.5 & 203.0 (141) \\
         & 0.8 & 0.5 & 40.3 (35) & 0.8 & 0.5 & 234.1 (141) \\
         & 0.6 & 0.5 & 49.0 (35) & 0.6 & 0.5 & 291.9 (141) \\
\hline
\end{tabular}
\end{center}
\caption{We fitted the spectra in the energy range 3--10keV using an
absorbed stratified accretion column model and stepped through a range
of white dwarf masses and metal abundance. We then fixed these
emission models and fitted in the extended energy range 0.2--10.0keV
after adding a blackbody and partial covering component. In comparison
when we allowed the parameters of the stratified emission to vary the
fits were \uchi=241.7 (237) and \uchi=190.4 (136) for V347 Pav and GG
Leo respectively. For five extra parameters the difference in
\uchi is 9.24 and 11.3 at the 90\% and 99\% confidence levels.}
\label{above3kev}
\end{table}

\subsection{The UV fluxes}

We use the best fit models in Table 2 to predict the flux in the OM UV
filters. The flux from the post-shock accretion flow falls well below
the observed UV flux in all three systems. This indicates that the
emission from the unheated photosphere of the white dwarf must
dominate in the UV.

We can estimate the temperature for our objects based on their UV
filter flux ratios. Here we use the mean flux taken from those orbital
phases where there were data from both filters. We use model white
dwarf Hydrogen atmosphere models of covering a range of temperature
which were kindly supplied by Detlev Koester.  We take the neutral
absorption reported from our fits to the X-ray data (Table \ref{fits})
and convert this to optical extinction using the relationship of
Predehl \& Schmitt (1995).  Assuming the absorption derived from the
X-ray fits are appropriate we find that the temperature of the white
dwarf us is $<$10200K for V347 Pav and 9200--10400K and 9200-11500K
for GG Leo and EU UMa respectively.  This is similar to the
temperatures determined for the unheated white dwarf by G\"{a}nsicke
(1998). (The temperatures determined using a simple blackbody
approximation are similar being $\sim$3000 K hotter than that derived
using white dwarf atmosphere models).

Using the white dwarf atmosphere models of Koester we can set the
normalisation so that it gives the measured flux in the UV filters
(this is in addition to the model used to fit the X-ray spectra). If
we assume a white dwarf of certain radius (based on the derived masses
in \S \ref{mass}) we can obtain a distance estimate. Naturally since
the luminosity of the white dwarf is proportional to $T^{4}$ and the
calibration of the UV filters is reliable to $\sim$5 percent, we
should approach these distances with due caution. With this in mind,
we estimate distances of $\sim$40-50pc, 50-70pc and 60-80pc for V347
Pav, GG Leo and EU UMa respectively.

\section{acknowledgments}

We thank Detlev Koester for kindly supplying his white dwarf model
atmosphere spectra. This paper is based on observations obtained with
XMM-Newton, an ESA science mission with instruments and contributions
directly funded by ESA Member States and the USA (NASA).

\end{document}